\documentclass[final]{ws-ijmpd}

\begin{document}

\markboth{R. Opher and A. Pelinson}{What is the SUSY mass scale?}

%%%%%%%%%%%%%%%%%%%%% Publisher's Area please ignore %%%%%%%%%%%%%%%
%
\catchline{}{}{}{}{}
%
%%%%%%%%%%%%%%%%%%%%%%%%%%%%%%%%%%%%%%%%%%%%%%%%%%%%%%%%%%%%%%%%%%%%

\title{What is the SUSY mass scale?}

\author{Reuven Opher{$^\ast$}  and Ana Pelinson{$^\dag$} }

\address{Instituto  de
Astronomia, Geof\'{\i}sica e Ci\^encias Atmosf\'ericas, Universidade
de S\~{a}o
Paulo\\ Rua do Mat\~{a}o, 1226 - Cidade Universit\'aria\\
CEP 05508-900 - S\~{a}o Paulo, S.P., Brazil\\
$\ast$opher@astro.iag.usp.br  \\
$\dag$anapel@astro.iag.usp.br}

\maketitle

\begin{history}
\received{Day Month Year}
\revised{Day Month Year}
\comby{Managing Editor}
\end{history}

\begin{abstract}
The energy, or mass scale $M_{\rm SUSY}$, of the supersymmetry
(SUSY) phase transition is, as yet, unknown. If it is very high
(i.e., $\gg 10^{3}\, {\rm GeV}$), terrestrial accelerators will not
be able to measure it. We determine $M_{\rm SUSY}$ here by combining
theory with the cosmic microwave background (CMB) data. Starobinsky
suggested an inflationary cosmological scenario in which inflation
is driven by quantum corrections to the vacuum Einstein's equation.
The modified Starobinsky model (MSM) is a natural extension of this.
In the MSM, the quantum corrections are the quantum fluctuations of
the supersymetric (SUSY) particles, whose particle content creates
inflation and whose masses terminate it. Since the MSM is difficult
to solve until the end of the inflation period, we assume here that
an effective inflaton potential (EIP) that reproduces the time
dependence of the cosmological scale factor of the MSM can be used
to make predictions for the MSM. We predict the SUSY mass scale to
be $M_{\rm SUSY}\simeq 10^{15}{\rm GeV}$, thus satisfying the
requirement that the predicted density fluctuations of the MSM be in
agreement with the observed CMB data.
\end{abstract}

\keywords{supersymmetry; inflation; density fluctuations.}

%%SECTION%I%%%%%%%%%%%%%%%%%%%%%%%%%%%%%%%%%%%%%%%%%%%%%%%%%%%%%%%%
\section{Introduction}

Cosmological models based on the anomaly-induced effective action of gravity take
into account the vacuum quantum effects of particles in the early universe. These
models naturally lead to inflation, which is a direct consequence of the assumption
that, at high energies, all particles can be described by massless, conformally
invariant fields with negligible interaction between them.
\cite{fhh}\cdash\cite{wave}

In the modified version of the original Starobinsky model
\cite{star1} [modified Starobinsky model (MSM)], it was assumed that
some of the scalar and fermion fields are massive.
\cite{anju}\cdash\cite{asta} If we also assume a supersymmetric
particle content, inflation is, then, stable during the period which
starts at a sub-Planck scale, continuing almost until the end of
inflation, when most of the sparticles decouple.
\cite{asta}\cdash\cite{gorb} Subsequently, the universe enters into
an unstable regime, eventually undergoing a transition to the FRW
evolution. It is to be noted that this occurs without any need for
fine-tuning, regardless of the values of the cosmological constant
$\Lambda$ and the curvature parameter $k$.

In fact both the cosmological constant $\Lambda$ and the curvature
$k$ are very small in the context of the high-T early universe.
Present observations indicate that the curvature $k$-term is less
than $10\%$ of the energy density term in the Friedmann equation and
that the $\Lambda$ term is comparable to the energy density term in
the $\Lambda$CDM model. Extrapolating back to the high-T early
universe, the $k$-term is then $\lesssim 10^{-50}\%$ and the
$\Lambda$ term $\lesssim 10^{-100}\%$ of the energy density term.
The $k$-term is thus not included in our equations and the
$\Lambda$-term is included up to Eq.(\ref{parabola}), just for
completeness. We set $\Lambda=0$ in the equations after
Eq.(\ref{parabola}).

It is not possible to solve analytically the equation for the cosmological scale
factor $a(t)$ in the MSM. However an approximate solution for $a(t)$ during
inflation was obtained and confirmed by numerical analysis with very high precision.
\cite{asta}

We assume here that an effective inflaton potential (EIP) that reproduces the time
dependence of $a(t)$ of the MSM during the inflationary period can be used to make
predictions of the MSM throughout this period, including its end. We use a reverse
engineering method, discussed in Ref.~\refcite{ellis1}, to derive the EIP from
$a(t)$.

In $\S$ 2, we give a brief review of anomaly-induced inflation of the MSM and the
approximate time dependence of $a(t)$. The effective potential and value for $M_{\rm
SUSY}$ are derived in $\S$ 3. Our conclusions are presented in $\S$ 4.

%%SECTION%II%%%%%%%%%%%%%%%%%%%%%%%%%%%%%%%%%%%%%%%%%%%%%%%%%%%%%%%%
\section{The approximate solution of the cosmological scale factor in the modified Starobinsky model}

The interesting property of the Starobinsky model is that inflation
is a result of vacuum quantum effects of predicted (e.g., SUSY)
particles. In the simplest case, based on the effects of massless
fields, the leading quantum phenomenon is the conformal anomaly. The
underlying theory includes $N_0$ scalars, $N_{1/2}$ Dirac spinors,
and $N_1$ vectors, corresponding to the particle content of the
quantum theory. They are not necessarily the real matter filling the
universe, whose energy density is assumed to be negligible during
inflation. The vacuum quantum effects originate from the virtual
particles.

Due to conformal invariance, the fields decouple from the conformal factor of the
metric. In this case, the dominating quantum effect is the trace anomaly, which
comes from the renormalization of the conformal invariant part of the vacuum action,
\cite{birdav,book}
%%%%%%%%%%%%%%%%%%%%%%%%%%%%%%%%%%%%
\begin{equation} S_{vacuum}\, =\, S_{HD}\, + \,S_{EH}\,, \label{vacuum}
\end{equation}
%%%%%%%%%%%%%%%%%%%%%%%%%%%%%%%%%%%%
where the first term contains higher derivatives of the metric,
%%%%%%%%%%%%%%%%%%%%%%%%%%%%%%%%%%%%
\begin{equation} S_{HD}\, =\, \int d^4x\sqrt{-g}\, \left\{ a_1 C^2 + a_2 E +
a_3 {\nabla^2} R  \right\}\,, \label{higher} \end{equation}
%%%%%%%%%%%%%%%%%%%%%%%%%%%%%%%%%%%%
and
%%%%%%%%%%%%%%%%%%%%%%%%%%%%%%%%%%%%
$$ S_{EH}\, =\, -\,\frac{1}{16\pi G}\,\int d^4x\sqrt{-g}\,(R +
2\Lambda)\, $$
%%%%%%%%%%%%%%%%%%%%%%%%%%%%%%%%%%%%
is the Einstein-Hilbert term, where $\,a_{1,2,3},\,G,\,$ and
$\,\Lambda\,$ are the parameters of the vacuum action. $C^2$ and $E$
are the square of the Weyl tensor and the integrand of the
Gauss-Bonnet term, respectively:
%%%%%%%%%%%%%%%%%%%%%%%%%%%%%%%%%%%%
$$
C^2=R_{\mu\nu\alpha\beta}^2 - 2R_{\alpha\beta}^2 + 1/3\,R^2 \,,
$$
$$
E = R_{\mu\nu\alpha\beta}R^{\mu\nu\alpha\beta} -4
\,R_{\alpha\beta}R^{\alpha\beta} + R^2\,.
$$
%%%%%%%%%%%%%%%%%%%%%%%%%%%%%%%%%%%%
%%%%%%%%%%%%%%%%%%%%%%%%%%%%%%%%%%%%
%%%%%%%%%%%%%%%%%%%%%%%%%%%%%%%%%%%%%%%%%%%%%%%%%%%%%%%%%%%%%%%
%%%%%%%%%%%%%%%%%%%%%%%%%%%%%%%%%%%%%%%%%%%%%
%%%%%%%%%%%%%%%%%%%%%%%%%%%%%%%%%%%%%%%%%%%%%%%%%%%%%%%%%%%%%%%
%%%%%%%%%%%%%%%%%%%%%%%%%%%%%%%%%%%%%%%%%%%%%

Quantum corrections to the Einstein equation,
%%%%%%%%%%%%%%%%%%%%%%%%%%%%%%%%%%%%
$$ R_{\mu\nu}\,-\,\frac12\,R\,g_{\mu\nu}\,=\, 8\pi G\,<T_{\mu\nu}>
\,-\,\Lambda \,,$$
%%%%%%%%%%%%%%%%%%%%%%%%%%%%%%%%%%%%
produce a non-trivial effect due to the anomalous trace of the energy-momentum
tensor. The anomaly-induced effective action can be found explicitly.
\cite{book}\cdash\cite{frts} The expression for the anomalous energy-momentum tensor
is
%%%%%%%%%%%%%%%%%%%%%%%%%%%%%%%%%%%%
\begin{equation}
T\,\,=\,\,<T_\mu^\mu>\,\,=\,\,- \frac{2}{\sqrt{-g}}\,g_{\mu\nu}
\frac{\delta { \Gamma}}{\delta g_{\mu\nu}}\,\,=\,\, - \,(wC^2 + bE +
c{\nabla^2} R)\,, \label{main equation}
\end{equation}
%%%%%%%%%%%%%%%%%%%%%%%%%%%%%%%%%%%%
%%%%%%%%%%%%%%%%%%%%%%%%%%%%%%%%%%%%
where $w,\,b$ and $c$ are the $\beta$-functions for the parameters
$a_1,\,a_2,\,a_3$ in Eq.(\ref{higher}), respectively:
%%%%%%%%%%%%%%%%%%%%%%%%%%%%%%%%%%%%
\begin{equation} w \,=\, \frac{1}{(4\pi)^2}\,\Big( \frac{N_0}{120} +
\frac{N_{1/2}}{20} + \frac{N_1}{10} \Big)\,, \label{w}
\end{equation}
%%%%%%%%%%%%%%%%%%%%%%%%%%%%%%%%%%%%
\begin{equation} b\,=\, -\,\frac{1}{(4\pi)^2}\,\Big( \frac{N_0}{360} +
\frac{11\,N_{1/2}}{360} + \frac{31\,N_1}{180}\Big)\,, \label{b}
\end{equation}
%%%%%%%%%%%%%%%%%%%%%%%%%%%%%%%%%%%%
\begin{equation} c \,=\, \frac{1}{(4\pi)^2}\,\Big( \frac{N_0}{180} +
\frac{N_{1/2}}{30} - \frac{N_1}{10}\Big) \, \label{c}
\end{equation}
%%%%%%%%%%%%%%%%%%%%%%%%%%%%%%%%%%%%
and $\Gamma$ is the quantum correction to the classical vacuum
action, taking into account the vacuum quantum effects of the matter
fields, which are free, massless, and conformally coupled to the
metric. In terms of the new variables, $\bar{g}_{\mu\nu}$ and
$\sigma\,(=\ln a(t))$, where $a(t)$ is the cosmic scale factor and
$g_{\mu\nu}=\bar{g}_{\mu\nu}\,.\, e^{2\sigma}$,  the quantum
correction to the classical vacuum action is
%%%%%%%%%%%%%%%%%%%%%%%%%%%%%%%%%%%%
$$
{\bar \Gamma} = S_c[{\bar g}_{\mu\nu}] + \int d^4 x\sqrt{-{\bar
g}}\,\{ w\sigma {\bar C}^2 + b\sigma ({\bar E}-\frac{2}{3} {\bar
{\Box}} {\bar R}) + 2b\sigma{\bar \Delta}\sigma\,\}
$$
\begin{equation}
 - \frac{3c+2b}{36}\,\int d^4 x\sqrt{-g}\,R^2\,,
\label{quantum}
\end{equation}
%%%%%%%%%%%%%%%%%%%%%%%%%%%%%%%%%%%%
where $\,S_c[{\bar g}_{\mu\nu}]=S_c[g_{\mu\nu}]\,$ is some unknown
functional of the metric. In general, there is no standard method
for deriving $S_c[{\bar g}_{\mu\nu}]$. If we consider an isotropic
and homogeneous metric, $g_{\mu\nu}={\bar g}_{\mu\nu}\cdot
a^2(\eta)$, where $\eta$ is the conformal time, the conformal
functional $S_c[{\bar g}_{\mu\nu}]$ is constant. In this case,
$\,S_c[{\bar g}_{\mu\nu}]$ does not depend on $a(\eta)$ and,
therefore, does not contribute to the equations of motion
[Eq.(\ref{quantum})] which is an exact one-loop quantum correction.

The total action with quantum corrections is
%%%%%%%%%%%%%%%%%%%%%%%%%%%%%%%%%%%%
\begin{equation} S_{total}= S_{vacuum}+\Gamma \,, \label{massless} \end{equation}
%%%%%%%%%%%%%%%%%%%%%%%%%%%%%%%%%%%%
which leads to the following equation of motion for $a(t)$:
%%%%%%%%%%%%%%%%%%%%%%%%%%%%%%%%%%%%
\begin{equation} \frac{{\stackrel{....}{a}}}{a}
+3\,\frac{{\stackrel{.}{a}}}{a} \frac{{\stackrel{...}{a}}}{a}
+\frac{{\stackrel{..}{a}}^{2}}{a^{2}} -\left( 5+\frac{4b}{c}\right)
\frac{{\stackrel{..}{a}}}{a} \frac{{\stackrel{.}{a}}^{2}}{a^2}
-\frac{M^2_{\rm Pl}}{c} \left( \frac{{\stackrel{..}{a}}}{a}+
\frac{{\stackrel{.}{a}}^{2}}{a^{2}} -\frac{2\Lambda }{3}\right)
\,=\,0\,, \label{foe} \end{equation}
%%%%%%%%%%%%%%%%%%%%%%%%%%%%%%%%%%%%
where $\,M_{\rm Pl}=1/\sqrt{8\pi G}=2.44\times 10^{18}\,{\rm GeV}$
is the reduced Planck mass. The dot over the variable denotes the
derivative with respect to the physical time $t$, which is related
to the conformal time $\eta$ by the relation $dt=a(\eta)d\eta$.

The last term in Eq.(\ref{foe}) is the trace of the standard
classical Einstein equation when equated to $(3p-\rho)/6 c$, where
$p\,(\rho)$ is the pressure (energy density) of the free particles
and $c$ is given by Eq.(\ref{c}). There are only vacuum fluctuations
(virtual particles) in Eq.(\ref{foe}) and no free particles. The
other terms (the quantum corrections) become negligible if we assume
that we have a FRW evolution at late times with $a(t)\sim t^{2/3}$
in a matter dominated era, by massive particles having negligible
pressure, for example. The equation of motion [Eq.(\ref{foe})] is,
then, equal to $-\rho_M^0/6 a^3c$, instead of being homogeneous. The
quantum correction terms have ${\stackrel{....}{a}},\,
{\stackrel{.}{a}}{\stackrel{...}{a}},\,{\stackrel{..}{a}}^2,\,{\stackrel{..}{a}}{\stackrel{.}{a}}^2$,
which are proportional to the fourth derivative of time of the
cosmic scale factor $a$, while the classical Einstein term is
proportional to ${\stackrel{..}{a}}$ and ${\stackrel{.}{a}}^2$, the
second derivative of time of $a$. We thus expect that the quantum
correction terms will not contribute appreciably when the cosmic
scale factor is varying slowly with time (e.g., $a\propto t^{2/3}$
in a matter dominated universe or $a\propto t^{1/2}$ in a radiation
dominated universe).

Using the FRW metric with $k=0$, the solution to Eq.(\ref{foe}) is
%%%%%%%%%%%%%%%%%%%%%%%%%%%%%%%%%%%%
\begin{equation}
a(t) \,=\, a_0 \cdot \exp(Ht)\,, \label{flat solution}
\end{equation}
%%%%%%%%%%%%%%%%%%%%%%%%%%%%%%%%%%%%
where $H$ has the form
%%%%%%%%%%%%%%%%%%%%%%%%%%%%%%%%%%%%
\begin{equation} H\,=\, \frac{1}{\sqrt{(-32\pi G) b}} \cdot \left(\,1\pm \,
\sqrt{1+\frac{(64\pi G) \,b\,\Lambda }{3}\, \,}\right)^{1/2}\,.
\label{H}
\end{equation}
%%%%%%%%%%%%%%%%%%%%%%%%%%%%%%%%%%%%
We do not consider solutions with negative $H$ or $k=\pm 1$. The general equation of
motion for $a(t)$ and solutions with $k=\pm 1$ can be found in Ref.~\refcite{asta}.
When $\Lambda=0$, the solution Eq.(\ref{foe}) is
%%%%%%%%%%%%%%%%%%%%%%%%%%%%%%%%%%%%
\begin{equation}
H\,=\, \frac{1}{\sqrt{(-16\pi G)\,b}}\equiv H_S \,,\label{starh}
\end{equation}
%%%%%%%%%%%%%%%%%%%%%%%%%%%%%%%%%%%%
which is the solution for $a(t)$ in the original Starobinsky model. \cite{star1}

%%%%%%%%%%%%%%%%%%%%%%%%%%%%
%%%%%%%%%%%%%%%%%%%%%%%%%%%%

The inflationary solution (\ref{flat solution}) is stable for $c>0$ and unstable for
$c<0$, \cite{star1} regardless of the cosmological constant. \cite{anju} According
to Eq.(\ref{c}), the condition $c>0$ requires
%%%%%%%%%%%%%%%%%%%%%%%%%%%%%%%%%%%%
\begin{equation}
N_1 \,<\,\frac13\,N_{1/2}\,+\,\frac{1}{18}\,N_{0}\,.
\label{condition}
\end{equation}
%%%%%%%%%%%%%%%%%%%%%%%%%%%%%%%%%%
This enables the construction of an attractive inflationary scenario.
\cite{gracexit} The universe could start in a stable phase, such that inflation
starts regardless of the initial data. The simplest way to provide stability in
Eq.(\ref{condition}) is to assume that supersymmetry exists in the high energy
region at the beginning of inflation and that it is broken at a lower energy since
the sparticles are heavy and decouple. \cite{asta,gracexit} During inflation, $H$
decreases due to the massive fields and, at some point, the loops of the sparticles
decouple and the matter content $N_{0,1/2,1}$ becomes modified. As a result, the
inequality sign in Eq.(\ref{condition}) changes and the universe enters into an
unstable inflation regime with an eventual transition to the FRW evolution.

The intermediate transition epoch between inflation and post-inflation is
characterized by vacuum quantum effects of both massive and massless fields.
\cite{asta,shocom} In this epoch, the conformal invariance of the actions is
violated by the masses and, therefore, we cannot use the conformal anomaly to derive
quantum corrections. However, the conformal description of the massive theory in the
framework of the cosmon model can be used  \cite{cosmon}\cdash\cite{dragon} (see
also Refs.~\refcite{sola89}--\refcite{tomb} for similar analyzes and applications of
the cosmon method). A detailed discussion of the effects of massive fields,
including quantum corrections, can be found in Ref.~\refcite{asta}.

A leading-log approximation for the effective action for the massive
fields,
%%%%%%%%%%%%%%%%%%%%%%%%%%%%%%%%%%%%
$$
\Gamma \,\,=\,\,S_{HD}+S_c[g_{\mu\nu}, M] \,+\, \int d^4
x\sqrt{-{\bar g}} \,\{w{\bar C}^2\sigma + b({\bar E} -\frac23 {\bar
\nabla}^2 {\bar R})\sigma + 2 b\,\sigma{\bar \Delta}\sigma \}\,-
$$$$
\,-\, \frac{3c+2b}{36}\,\int d^4x\sqrt{-g}\,R^2\,-\,\int d^4
x\sqrt{-{\bar g}} \,e^{2\sigma}\,[{\bar R}+6({\bar \nabla}\sigma)^2]
\,\cdot\,\Big[\, \frac{1}{16\pi G} - f\cdot\sigma\,\Big]\,-
$$
%%%%%%%%%%%%%%%%%%%%%%%%%%%%%%%%%%%%
\begin{equation}
 - \int d^4 x\sqrt{-{\bar
g}}\,e^{4\sigma}\,\cdot\, \Big[\frac{\Lambda}{8\pi
G}\,-\,g\cdot\sigma\,\Big] \,, \label{quantum for massive}
\end{equation}
%%%%%%%%%%%%%%%%%%%%%%%%%%%%%%%%%%%%
holds in the high energy region until a ``cut off" scale, defined
such that at $H =M_*$ is reached, when a number of the sparticles
decouple. The inflation then becomes unstable, and enters the FRW
phase.

The equation of motion for $\sigma(t)=\ln {a(t)}$ in the flat case
$k=0$, has the form
$$ {\stackrel{....}{\sigma }}+7{\stackrel{...}{\sigma }}{\stackrel{.}{\sigma }}
+4\,{{\stackrel{..}{\sigma }}}^{2} +4\,\Big( 3-\,\frac{b}{c}\Big)
\,{\stackrel{..}{\sigma }}{{\stackrel{.}{\sigma }}}^{2}
-4\,\frac{b}{c}\,{{\stackrel{.}{\sigma }}}^{4} \,\, + \,\,
\frac{2\,\Lambda}{3}\frac{M^2_{\rm Pl}}{c} \,(1 -
\tilde{g}\sigma-\tilde{g}/4) \,-
$$
\begin{equation} - \,\frac{M^2_{\rm Pl}}{c}\,\Big[ \,(
{\stackrel{..}{\sigma}}\,+2{{\stackrel{.}{\sigma }}}^{2} )\cdot
(1-\tilde{f}\sigma)-\frac12\,\tilde{f}
\dot{\sigma}^2\,\Big]\,=\,0\,, \label{central sigma}
\end{equation} where
%%%%%%%%%%%%%%%%%%%%%%%%%%%%%%%%%%%%
$$ \tilde{f} = (16\pi G)\, f = \frac{2}{3{(4 \pi)}^
2}\,\sum_{f}\,\frac{N_f\,m_f^2}{M^2_{\rm Pl}}\,,$$
%%%%%%%%%%%%%%%%%%%%%%%%%%%%%%%%%%%%
\begin{equation}
\tilde{g} = \frac{g }{\Lambda/(8\pi G)}\,
\,=\,\frac{1}{2(4\pi)^2}\,\sum_{s}\,\frac{N_s\,m_s^4}{M^2_{\rm
Pl}\Lambda}
-\frac{2}{(4\pi)^2}\,\sum_{f}\,\frac{N_f\,m_f^4}{M^2_{\rm
Pl}\Lambda}\,. \label{replace11}
\end{equation}
%%%%%%%%%%%%%%%%%%%%%%%%%%%%%%%%%%%%
In Eq.(\ref{replace11}), the sums are taken over all fermions with a mass $m_f$ and
multiplicity $N_f$ as well as over all scalars with a mass $m_s$ and multiplicity
$N_s$. It can be seen that the higher-derivative terms of (\ref{quantum for
massive}) are identical to those for the massless fields, as expected.
\cite{mamo,birdav}

Although the solution of Eq.(\ref{central sigma}) can not be performed analytically,
an approximate solution can be obtained. From Eqs.(\ref{flat solution}) and
(\ref{H}), $\sigma(t)\left(=\ln a(t)\right)$ is proportional to $t$, for small $t$.
However, inflation slows down at larger $t$, when $\sigma(t)$ is no longer linear in
$t$ due to the masses of the particles. We then approximate $\sigma( t)$ as a second
order term in time, $\sigma(t)= A\times t + B\times t^2$, where $A$ and $B$ are
constants and $B$ is negative.
%%%%%%%%%%%%%%%%%%%%%%%%%%%%
The equation of motion, modified by the contribution of the massive
fields, produces the following approximate solution to
Eq.(\ref{central sigma}):
%%%%%%%%%%%%%%%%%%%%%%%%%%%%%%%%%%%
\begin{equation} \sigma(t)\,=\,\frac{H_{a}}{\sqrt{8\pi G}} \,t\,-\frac{1}{(8\pi
G)}\frac{H_{a}^2\tilde{f}}{4}\,\,t^2\,, \label{parabola}
\end{equation}
%%%%%%%%%%%%%%%%%%%%%%%%%%%%%%%%%%%%
%%%%%%%%%%%%%%%%%%%%%%%%%%%%%%%%%%%%
where $H_{a}= H_S/\,M_{\rm Pl}=1/\sqrt{-2\,b}\,$ is a dimensionless Hubble
parameter, which depends on the particle content of Eq.(\ref{b}). We use the
particle content $N_{1,1/2,0}=(12,48,104)$ of the minimum supersymmetric standard
model (MSSM) since it provides a stable inflation in the MSM and $\Lambda=0$.
\cite{asta}

%%%%%%%%%%%%%%%%%%%%%%%%%%%%%%%%%%%%%%%%%%%%%%%%FIG1
\begin{figure*}[h]
\resizebox{\columnwidth}{!}{\includegraphics{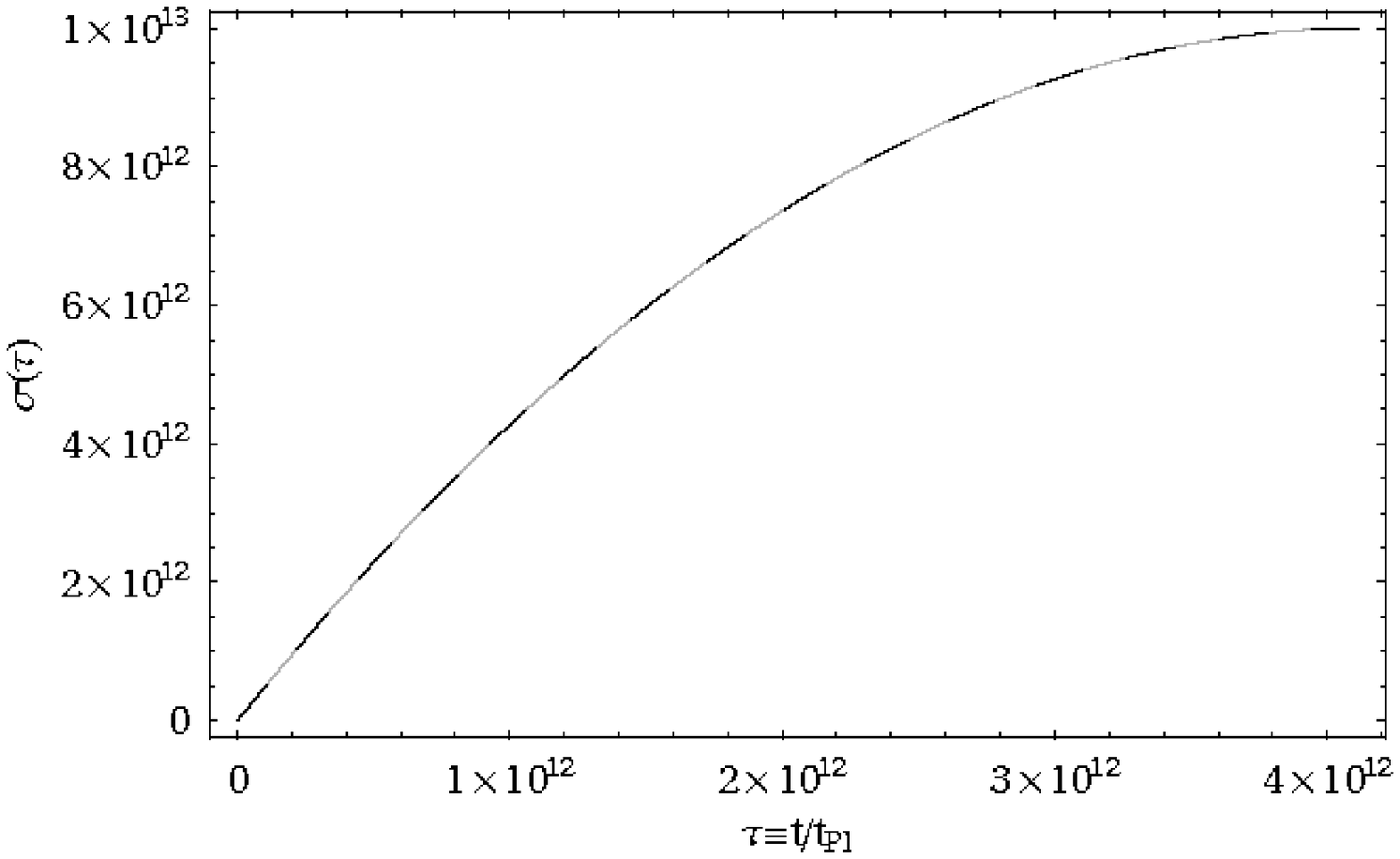}}
\resizebox{\columnwidth}{!}{\includegraphics{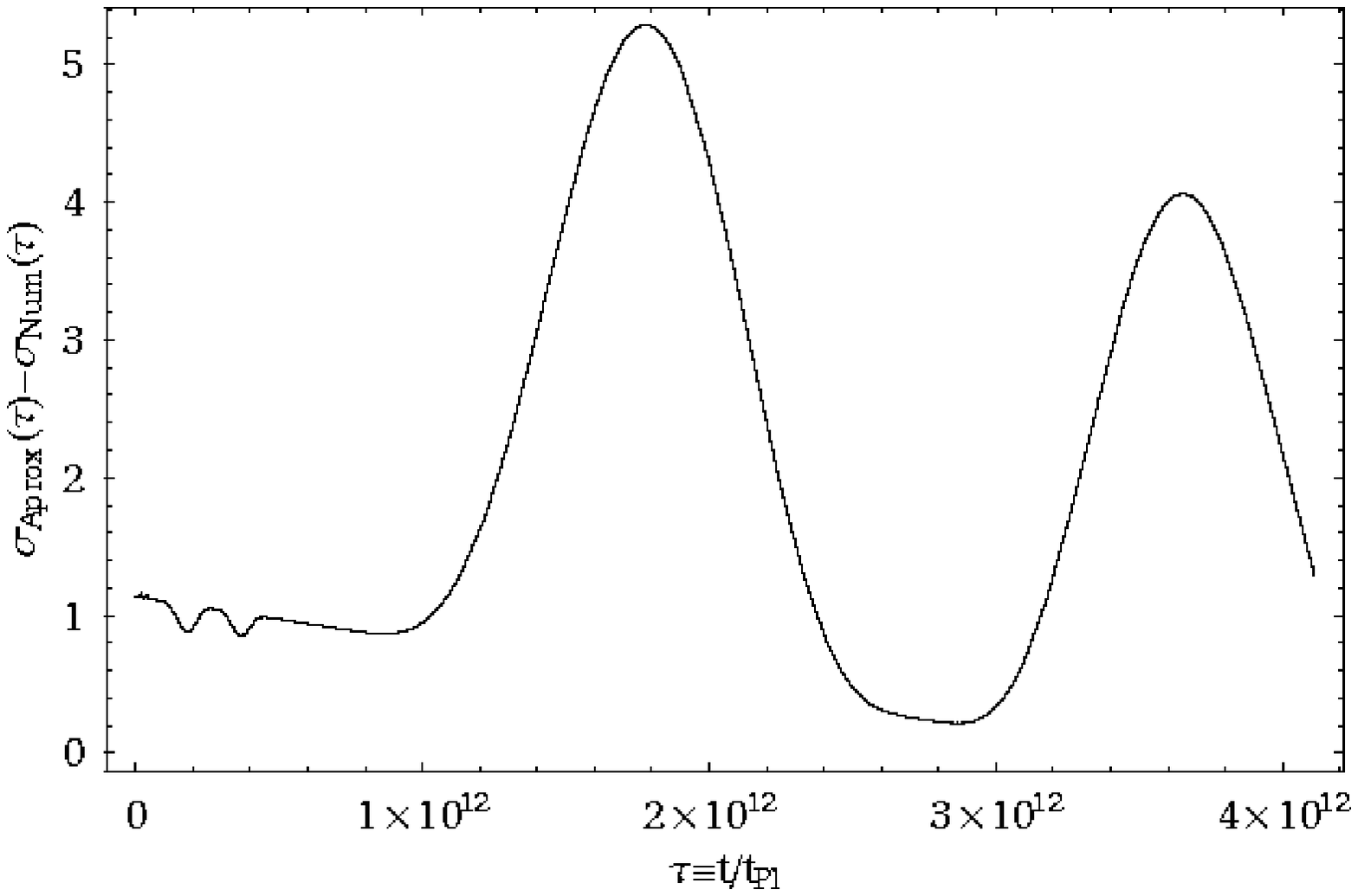}}
\caption{a) Top: The dashed line shows the numerical solution of
$\sigma(\tau\equiv t/\sqrt{8\pi}\,t_{\rm Pl})$, using
Eq.(\ref{central sigma}), for $\Lambda=0$, MSSM particle content and
$\tilde{f}\simeq 10^{-13}$. The gray line shows the parabolic
solution [Eq.(\ref{parabola})]. b) Bottom: The difference between
the approximate $\sigma_{\rm Aprox}$ [Eq.(\ref{parabola})] and the
numerical solution $\sigma_{\rm Num}$ [Eq.(\ref{central sigma})].
The difference corresponds to approximately one part in $10^{13}$.}
\label{sigcomp}
\end{figure*}
%%%%%%%%%%%%%%%%%%%%%%%%%%%%%%%%%%%%%%%%%%%%%%%%

In a previous paper, a relatively large value for $\tilde
f$($=10^{-6}$) was used as a strong test of the accuracy of the
approximate solution Eq.(\ref{parabola}). \cite{asta} The peak value
of Eq.(\ref{parabola}) with this value of $\tilde f$ was compared
with the peak value of the numerical solution of Eq.(\ref{central
sigma}) and found to differ by only one part in $10^6$. In
Fig.~\ref{sigcomp} we plot $\sigma(\tau)$ vs $\tau(\equiv
t/\sqrt{8\pi}\,t_{\rm Pl})$, where $t_{\rm Pl}=1/\sqrt{8\pi}M_{\rm
Pl} \simeq 5.3\times 10^{-44} {\rm sec}$ is the Planck time, for
$\tilde{f}\simeq 10^{-13}$ [Eq.(\ref{fresult})] for the numerical
solution [Eq.(\ref{central sigma})] and for the approximate
parabolic solution [Eq.(\ref{parabola})]. The approximate parabolic
and numerical solutions basically coincide. The two curves differ by
one part in $10^{13}$, as seen in Fig.~\ref{sigcomp}b. Thus, for any
value of $\tilde f$ on the order of $10^{-6}$ or less, the
approximate solution is extremely accurate. The numerical analysis
confirms the parabolic dependence of Eq.(\ref{parabola}) to a very
high precision up to $H\sim M_{\ast}$, where $M_{\ast}$ is the scale
when most of the sparticles decouple, the inequality in
Eq.(\ref{condition}) changes sign, and inflation becomes unstable.

In $\S$ 3, we derive the reverse engineered inflaton potential that produces the
approximate solution [Eq.(\ref{parabola})]. In the next section, we present the
theory and method for constructing the potential. \cite{ellis1}

%%%%%%%%%%%%%%%%%%%%%%%%%%%%
%%%%%%%%%%%%%%%%%%%%%%%%%%%%

%%SECTION%III%%%%%%%%%%%%%%%%%%%%%%%%%%%%%%%%%%%%%%%%%%%%%%%%%%%%%%
\section{The effective potential of the modified Starobinsky model}

Consider the dynamics of a Robertson Walker universe model with a classical scalar
field $\phi(t)$, the inflaton and a non-interacting fluid. Following
Ref.~\refcite{ellis1}, we construct the potential $V(\phi)$ from the set of
equations
%%%%%%%%%%%%%%%%%%%%%%%%%%%%%%%%%%%%
\begin{equation} V(\phi (t))=\frac{1}{(8\pi G)}\left( \dot{H}+3H^{2}\right)
\label{poteq} \end{equation}
%%%%%%%%%%%%%%%%%%%%%%%%%%%%%%%%%%%%
and
%%%%%%%%%%%%%%%%%%%%%%%%%%%%%%%%%%%%
\begin{equation} \dot{\phi}^{2}=-\frac{1}{(4\pi G)}\dot{H} \,. \label{phieq}
\end{equation}
%%%%%%%%%%%%%%%%%%%%%%%%%%%%%%%%%%%%
Using the above equations, we construct the potential of the MSM:
From Eq.(\ref{parabola}), the Hubble parameter is
%%%%%%%%%%%%%%%%%%%%%%%%%%%%%%%%%%%%
\begin{equation}
H(t)=\frac{H_{a}}{\sqrt{8\pi G}} - \frac{1}{(8\pi
G)}\frac{H_{a}^2\tilde{f}}{2}\,t\,. \label{Hdet} \end{equation}
%%%%%%%%%%%%%%%%%%%%%%%%%%%%%%%%%%%%
Substituting Eq.(\ref{Hdet}) into Eq.(\ref{phieq}), we obtain
%%%%%%%%%%%%%%%%%%%%%%%%%%%%%%%%%%%%
\begin{equation} t(\phi)=\pm \frac{(8\pi
G)}{H_{a}\sqrt{\tilde{f}}}\left(\phi(t)-\phi_{0}\right),\,
\label{phidet} \end{equation}
%%%%%%%%%%%%%%%%%%%%%%%%%%%%%%%%%%%%
where $\left| \phi_0 \right|>\left| \phi \right|$. Choosing the positive sign in
Eq.(\ref{phidet}), we have $-\infty < \phi< 0$. \cite{ellis1} Substituting
Eq.(\ref{phidet}) into Eq.(\ref{Hdet}),
%%%%%%%%%%%%%%%%%%%%%%%%%%%%%%%%%%%%
\begin{equation} H(\phi)=\frac{H_{a}}{\sqrt{8\pi G}} -
\frac{H_{a}\sqrt{\tilde{f}}}{2}\left( \phi(t)-\phi_0\right) \,.
\label{Hphi} \end{equation}
%%%%%%%%%%%%%%%%%%%%%%%%%%%%%%%%%%%%
We can find $V(\phi)$ by substituting Eq.(\ref{Hphi}) into
Eq.(\ref{poteq}) or $V(t)$ substituting Eq.(\ref{Hdet}) into
Eq.(\ref{poteq}):
%%%%%%%%%%%%%%%%%%%%%%%%%%%%%%%%%%%%
\begin{equation}
V(t)=\frac{1}{(8\pi G)}\left[-\frac{{H^2_a \tilde{f}}}{2(8\pi
G)}+3\left(\frac{H_a}{\sqrt{8\pi G}}-\frac{{H^2_a\tilde{f}}}{2(8\pi
G)}\,t\right)^2\right]\,. \label{vdet}
\end{equation}
%%%%%%%%%%%%%%%%%%%%%%%%%%%%%%%%%%%%

\begin{figure*}
\resizebox{\columnwidth}{!}{\includegraphics{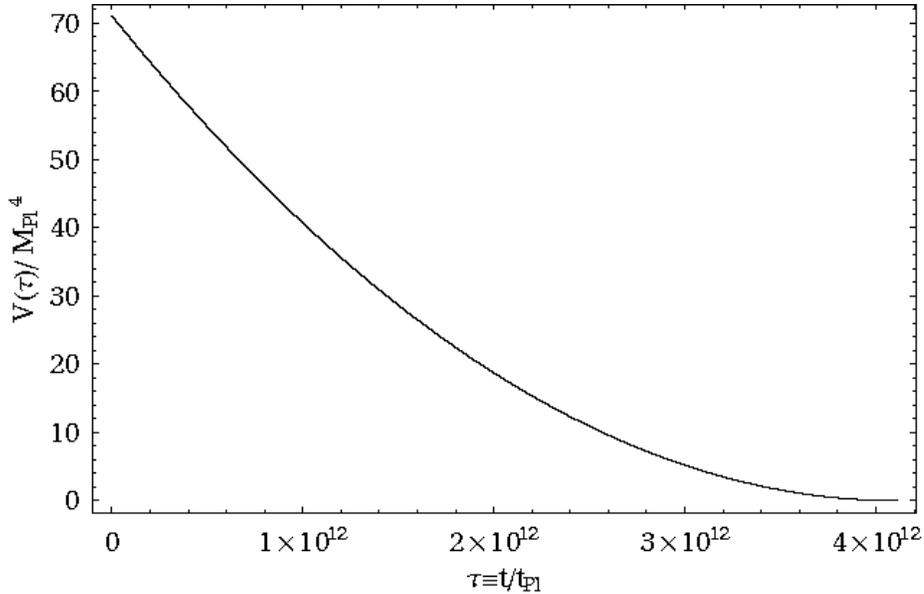}}
\caption{The potential [Eq.(\ref{vdet})] as a function of
$\tau(\equiv t/\sqrt{8\pi}\,t_{\rm Pl})$ for $\tilde{f}\simeq
10^{-13}$ and the MSSM particle content.} \label{potdet}
\end{figure*}

In Fig.~\ref{potdet}, we show the effective potential until the end
of inflation as a function of $\tau\equiv t/\sqrt{8\pi}\,t_{\rm
Pl}$, using $\tilde{f}\simeq 10^{-13}$ [Eq.(\ref{fresult})] and the
MSSM particle content. The potential is negligible at the end of
inflation ($V_{\rm end}/M^4_{\rm Pl}\simeq {H_a^2}\tilde{f}$).

%%%%%%%%%%%%%%%%%%%%%%%%%%%%%%%%%%%%%%%%%%%%%%%%FIG2
%%%%%%%%%%%%%%%%%%%%%%%%%%%%%%%%%%%%%%%%%%%%%%%%

The end of inflation occurs near the maximum of $\sigma(t)$, when
$a(t)$ is no longer increasing exponentially with time at the mass
(energy) scale $M_{\ast}$:
%%%%%%%%%%%%%%%%%%%%%%%%%%%%%%%%%%%%
\begin{equation} \dot{\sigma}(t){\vert}_{t=t_{\rm end}}=M_{\ast}\,.\end{equation}
%%%%%%%%%%%%%%%%%%%%%%%%%%%%%%%%%%%%
We define the dimensionless parameter
%%%%%%%%%%%%%%%%%%%%%%%%%%%%%%%%%%%%
\begin{equation}
\mu=M_{\ast}{\sqrt{8\pi G}}\,. \label{massend}
\end{equation}
%%%%%%%%%%%%%%%%%%%%%%%%%%%%%%%%%%%%
The time as a function of $\mu$ at the end of inflation is
%%%%%%%%%%%%%%%%%%%%%%%%%%%%%%%%%%%%
\begin{equation} t_{\rm end}=\frac{2\sqrt{8\pi G}}{H_a
\tilde{f}}\left[1-\left(\frac{\mu}{H_a}\right)\right]
\,.\label{tend}\end{equation}
%%%%%%%%%%%%%%%%%%%%%%%%%%%%%%%%%%%%
Substituting $t_{\rm end}$ into Eq.(\ref{vdet}), we have
%%%%%%%%%%%%%%%%%%%%%%%%%%%%%%%%%%%%
$$V(t=t_{\rm end})/M^4_{\rm Pl}={H_a^2}\tilde{f}\,,$$
%%%%%%%%%%%%%%%%%%%%%%%%%%%%%%%%%%%%
where $M_{\rm Pl}^2=1/(8\pi G)$. Since from Eq.(\ref{fresult})
$\tilde{f}$ is small, $V(\phi)$ is small at the end of inflation,
compared with its initial value.

The slow roll parameters in terms of the Hubble parameter are
\cite{liddle}
%%%%%%%%%%%%%%%%%%%%%%%%%%%%%%%%%%%%
$$\epsilon \equiv 2\,M_{\rm Pl}\left[
\frac{H^{\,\prime }\left( \phi \right) }{H\left( \phi \right)
}\right] ^{2}\,, $$
%%%%%%%%%%%%%%%%%%%%%%%%%%%%%%%%%%%%
%%%%%%%%%%%%%%%%%%%%%%%%%%%%%%%%%%%%
\begin{equation}
\eta \equiv 2\,M_{\rm Pl}\left[ \frac{H^{\,\prime \prime }\left(
\phi \right) }{H\left( \phi \right) }\right]\,. \label{sroll}
\end{equation}
%%%%%%%%%%%%%%%%%%%%%%%%%%%%%%%%%%%%

The value for $\mu$, that characterizes the end of inflation is
determined by the condition that the slow-roll parameter of
inflation is $\epsilon= 1$. Using this condition and
Eqs.(\ref{Hphi}) and (\ref{sroll}), we obtain
%%%%%%%%%%%%%%%%%%%%%%%%%%%%%%%%%%%%
\begin{equation} \mu\approx H_a\sqrt{\frac{\tilde{f}}{2}}\,. \label{mast} \end{equation}
%%%%%%%%%%%%%%%%%%%%%%%%%%%%%%%%%%%%

The number of $e$-folds of inflation before $t=t_{\rm end}$ is
%%%%%%%%%%%%%%%%%%%%%%%%%%%%%%%%%%%%
\begin{equation}
N=\int_{t_{60}}^{t_{\rm end}}H(t)dt=\sigma(t_{\rm
end})-\sigma(t_{60})\,. \label{nef}
\end{equation}
%%%%%%%%%%%%%%%%%%%%%%%%%%%%%%%%%%%%

We are interested in $N\simeq 60$, the approximate time $t_{60}$
when the observed density (scalar) fluctuations and the primordial
gravitational (tensor) fluctuations from inflation were created.
Substituting $t_{\rm end}$ in Eq.(\ref{parabola}), we find
$\sigma_{\rm end}$. From (\ref{nef}), we get $\sigma_{60}$. Solving
(\ref{parabola}) for $t_{60}$, we obtain
%%%%%%%%%%%%%%%%%%%%%%%%%%%%%%%%%%%%
\begin{equation} t_{60}=\frac{2\sqrt{8\pi G}}{H_a
\tilde{f}}\left[1-\sqrt{\left(\frac{\mu}{H_a}\right)^2+60\tilde{f}}\right]
\,.\label{t50}\end{equation}
%%%%%%%%%%%%%%%%%%%%%%%%%%%%%%%%%%%%
Substituting $t_{60}$ from Eq.(\ref{t50}) and $\mu$ from
Eq.(\ref{mast}) into Eq.(\ref{vdet}), we obtain the density
fluctuations,
%%%%%%%%%%%%%%%%%%%%%%%%%%%%%%%%%%%%
\begin{equation}
 \frac{\delta \rho }{\rho }=2\,M_{\rm Pl}\,\frac{V^{{3}/{2}}(t)}{V^{\prime}(t){dt}/{d\phi}} \vert_{t=t_{60}} \approx
5.4 H_a\sqrt{{\tilde{f}}/{2}} \label{dflut2} \end{equation}
%%%%%%%%%%%%%%%%%%%%%%%%%%%%%%%%%%%%
that is observed to be $ \approx10^{-5}$. From the requirement that
$ {\delta \rho }/{\rho }\simeq 10^{-5}$, we find that
%%%%%%%%%%%%%%%%%%%%%%%%%%%%%%%%%%%%
\begin{equation}
\tilde{f}\approx 2.05 \times10^{-13}\,. \label{fresult}
\end{equation}
%%%%%%%%%%%%%%%%%%%%%%%%%%%%%%%%%%%%
Using this result in Eqs.(\ref{mast}) and (\ref{massend}), we obtain the scale
%%%%%%%%%%%%%%%%%%%%%%%%%%%%%%%%%%%%
\begin{equation} M_{\ast}\approx 4.5\times
10^{12}{\rm GeV}\,,\label{SUSY}\end{equation}
%%%%%%%%%%%%%%%%%%%%%%%%%%%%%%%%%%%%
which is the value of the Hubble rate $H_{\ast}$ when the massive
particles decouple. Particles decouple when their masses are of the
order of the temperature or of order $H_{\ast}^{1/2}M_{\rm
Pl}^{1/2}$, which gives
\begin{equation}
M_{\rm SUSY}\simeq \sqrt{M_{\ast} M_{\rm Pl}}\simeq 10^{15} {\rm
GeV}\,. \label{SUSYSCALE}
\end{equation}
%%%%%%%%%%%%%%%%%%%%%%%%%%%%%%%%%%%%
This value is greater than the electroweak scale ($\sim 10^3{\rm
GeV}$) and consistent with the predicted GUT scale ($\sim
10^{14}\,-\,10^{16}{\rm GeV}$).

%%SECTION%IV%%%%%%%%%%%%%%%%%%%%%%%%%%%%%%%%%%%%%%%%%%%%%%%%%%%%%%%%
\section{Conclusions}

A scalar potential was constructed using the approximate solution
for the time dependence of the cosmological scale factor of the MSM
during inflation. The potential was normalized at a time $\sim 60$
$e$-folds before the end of inflation in order to obtain the
observed level of density fluctuations in the CMB,
$\delta\rho/\rho\sim 10^{-5}$. The mass (energy) scale of the MSM at
the end of the inflation, $M_{\ast}\simeq 10^{12}{\rm GeV}$, which
we identify with the Hubble rate when the massive particles
decouple, predicts a SUSY scale $M_{\rm SUSY}$, consistent with the
GUT scale $M_{\rm SUSY}\simeq 10^{15} {\rm GeV}$.

\section*{Acknowledgments}

The authors thank I.L. Shapiro and the anonymous referee for helpful
comments. R.O. thanks the Brazilian agencies FAPESP (00/06770-2) and
CNPq (300414/82-0) for partial support. A.P. thanks FAPESP for
financial support (03/04516-0 and 00/06770-2).

%\section{References}

%\begin{thebibliography}{000} %for 3 digits

\end{document}